\documentstyle[prl,twocolumn,aps,floats]{revtex} 
\input{epsf} 
 
\begin{document} 
\draft 
\twocolumn 
\title{Bouncing--ball tunneling in quantum dots} 
 
\author{Gregor Hackenbroich$^1$ and Rafael A.\ Mendez$^{1,2}$} 
\address{$^1$Universit\"at GH Essen, Fachbereich 7, 45117 Essen, 
Germany} 
\address{$^2$Centro de Ciencias F\'{\i}sicas UNAM, Cuernavaca, Morelos,
Mexico}
 
\date{ \today} 
 
\maketitle 
 
\begin{abstract} 
  We show that tunneling through quantum dots can be completely
  dominated by states quantized on stable bouncing--ball orbits.
  The fingerprints of bouncing--ball tunneling are sequences of
  Coulomb blockade peaks with strongly correlated peak height and
  asymmetric peak line shape. Our results are in agreement with the
  striking correlations of peak height and transmission phase found
  in recent interference experiments with quantum dots.

\end{abstract} 
\vspace*{-0.05 truein} 

Transport experiments with semiconductor quantum dots yield
important information on the energy levels and the wave functions of
confined many--electron systems \cite{Kou97}. In the Coulomb
blockade regime current can flow only if two different charge states
of the quantum dot are tuned to have the same energy. This produces
large nearly equally spaced conductance peaks as a function of the
dot potential (tuned via electrostatic gates). At low temperatures,
the magnitude of each conductance peak directly measures the
magnitude of a single resonant wave function near the contacts to
the leads. Some years ago, a statistical theory for the peak height
distribution was developed \cite{Jal92} based on the assumption that
the wave functions are completely uncorrelated and described by
random matrix theory. Recent experiments \cite{Cha96,Fol96} found
good agreement with the predicted distributions. At the same time,
one of these experiments \cite{Fol96} reported large correlation
between the heights of adjacent peaks. Much stronger correlations of
{\em both} the peak height and the phase of the transmission
amplitude through a quantum dot were observed in novel interference
experiments \cite{Yac95,Schu97}. The correlations in these
experiments comprised {\em all} Coulomb blockade peaks found in a
gate voltage scan, yielding sequences of more than 10 correlated
peaks.
 
Theoretical work \cite{Alh98} demonstrated that nonzero temperature
can partially account for but not fully explain the peak
correlations observed in Ref.~\cite{Fol96}. More recently
\cite{Nar99}, it was shown that additional correlations can arise
from short periodic orbits. Such orbits lead to deviations from
random matrix predictions even for dots whose classical dynamics is
completely chaotic. The predicted peak modulations \cite{Nar99} are
of the type observed in Ref.~\cite{Fol96}, but can not account for
the much stronger correlations in the interference experiments
\cite{Yac95,Schu97}. Despite of considerable theoretical efforts
\cite{Hac97,Sil99,Bal99} the origin of these correlations has not 
been understood. In particular, the most striking feature of the
experiments \cite{Yac95,Schu97}, that the correlations comprise {\em
all} observed Coulomb peaks, has not been explained. We address this
feature below.

We show that tunneling through quantum dots can be completely
dominated by wave functions quantized on stable bouncing--ball
orbits. Bouncing--ball tunneling (BBT) dominates the transport if
(i) the leads are attached opposite to each other and (ii) if they
are sufficiently wide providing near normal injection of electrons into
the dot. Under the conditions (i) and (ii) the dot conductance
exhibits sequences of strongly correlated Coulomb peaks.  All peaks
within a sequence are dominated by the same bouncing--ball mode. The
{\em unique fingerprint} of BBT is a characteristic asymmetry of the
peak line shape for the peaks in the tails of the sequences. This
asymmetry is caused by the breaking of the particle--hole symmetry
as the bouncing--ball state moves away from the Fermi energy. We
find the asymmetry in the line shapes measured in
Ref.~\cite{Schu97}, and suggest that the inspection of line shapes
be used as a diagnostic for BBT in other experiments on quantum
dots.

Our starting point is the transmission probability $T_{\rm ring}$
through an Aharonov--Bohm ring with a quantum dot embedded in one
arm. We assume that the dot is in the Coulomb blockade regime. The
tunnel barriers around the quantum dot suppress multiple reflections
of electrons across the ring, and $T_{\rm ring}$ is given by
\cite{Bal99}
\begin{eqnarray} 
\label{eq_trans}
T_{\rm ring} = \int dE
\left( - {\partial f \over \partial E} \right) \left| t_0 + t(E) 
\exp[2 \pi i \Phi / \Phi_0] \right|^2 . 
\end{eqnarray}
Here, $\Phi$ is the magnetic flux through the ring, $\Phi_0$ is the
flux quantum, and $t_0$, $t(E)$ are the transmission amplitudes
through the free arm and the arm with the quantum dot, respectively
($t_0$ is assumed to be independent of injection energy). The ring
is connected to two reservoirs labeled up (U) and down (D) occupied
according to the Fermi distribution $f$. Below, we investigate the
phase $\phi_{\rm QD} \equiv {\rm arg} \int dE ( -\partial f /
\partial E ) t(E)$ and the transmission coefficient $T_{\rm QD}
\equiv \int dE (- \partial f / \partial E ) |t(E)|^2$.  Both can be
measured in quantum dot interference experiments
\cite{Yac95,Schu97}.

The transmission amplitude $t(E)$ is related to the retarded Green
function $G_{pq}$ of the quantum dot,
\begin{eqnarray}
\label{ampliude}
t(E) = \sum_{pq} V_p^U V_q^D G_{pq}^{ }(E) ,
\end{eqnarray}
where $V_p^{l}$ is the matrix element for tunneling between level p
in the dot and the reservoir $l=U,D$. For a weakly coupled dot the
width of the level $p$ is given by $\Gamma_p = \Gamma_p^U +
\Gamma_p^D$, where $\Gamma_p^{l}=|V_p^{l}|^2$ is the partial width
for decay into the reservoir $l$. The matrix element is expressed
\cite{Nar99} as
\begin{eqnarray}
\label{eq_over}
V_p^l = \left( { \hbar^2 \over m^*} \right)   
\left. \int d s \psi_l (s,z) \partial_z \psi_p^*(s,z) 
\right|_{z=0} ,
\end{eqnarray}
where the integration is performed along the edge between the
potential barrier and the quantum dot ($\partial_z$ denotes the
derivative normal to the barrier). The wave function $\psi_p$
corresponds to Dirichlet boundary conditions in the dot, while the
barrier tunneling is fully included in the lead wave function
$\psi_l$. The transverse potential in the tunneling region can be
taken quadratic \cite{Nar99} yielding $\psi_l \sim c_l
\exp[-(s-s_l)^2/2 a_{\rm eff}^2]$, where $s$ is the transverse
coordinate, $s_l$ the center of the constriction and $a_{\rm eff}$ its
effective width. We can restrict ourselves to the lowest transverse
mode since higher modes are suppressed by the barrier penetration
factor (included in $c_l$). The Gaussian form of the lead wave
function is convenient but not crucial for the results presented below.

We obtain the dot wave functions $\psi_p$ for an effective potential
which accounts for the dot confining potential and the interactions
in the dot in a mean--field sense. We use a billiard approximation
and model the dot by a hard--wall potential at the boundary,
parameterized in polar coordinates by $R(\phi) = R_0 [1+\epsilon
\cos (2 \phi)]$. The parameter $\epsilon$ measures the quadrupolar
deformation out of circular shape. We use nonzero $\epsilon$ to
mimic \cite{shape} the dots used in \cite{Yac95,Schu97}. As in the
experiments, we assume that the leads are attached opposite to each
other at the boundary points closest to the origin (the points with
$\phi = \pm \pi/2$). In contrast to most studies of quantum dots
which assume the dynamics in the dot to be either regular or fully
chaotic, our boundary parameterization yields mixed classical motion.
This is illustrated for $\epsilon=0.2$ in Fig.~2, showing chaotic
motion in most of phase space {\em except} for two large islands
associated with stable bouncing--ball motion between the contacts to
the leads.  We note that billiards of quadrupolar or similar shape
have recently been used in studies of optical microresonators
\cite{Noe97,NarHac99}.

\begin{figure}[t] 
\begin{center} 
\leavevmode 
\vspace*{-1.1cm} 
\epsfxsize = 7.2cm 
\epsfbox{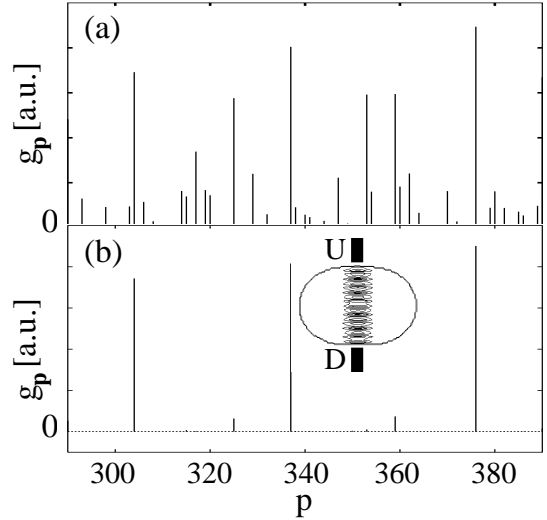} 
\end{center}
\vspace*{0.9cm} 
\caption{Coupling strength $g_p$ for a sequence of 100 states in the
  interval $290 \le p \le 390$. Results are for a quantum dot with
  quadrupolar shape and deformation $\epsilon=0.2$. (a) Narrow
  leads: $k a_{\rm eff} =0.1$. (b) Wide leads: $k a_{\rm eff}=5.0$.
  Here $k=(2 m^* E_p)^{1/2}/\hbar$ is the wave number associated
  with the bouncing--ball state $p=337$. Inset: Real--space
  projection of the state $p=337$.}
\end{figure}

In Fig.~1 we show the coupling strength $g_p \equiv \Gamma_p^U
\Gamma_p^D/(\Gamma_p^U +\Gamma_p^D)$ over a sequence of 100 billiard
states for two different values of $a_{\em eff}$. Note that $g_p$ is
proportional to the conductance peak height $G_p =(e^2/h) (\pi /2
kT) g_p$ measured in low temperature Coulomb blockade experiments
\cite{Kou97}. We calculated $g_p$ from Eq.~(\ref{eq_over}) using
quantum dot wave functions obtained numerically with the boundary
integral method \cite{BouInt}.  For narrow leads the peak heights
fluctuate with the level index $p$, occasionally one observes
systematic peak modulations as reported in Ref.~\cite{Nar99}. In
striking contrast, the results for wide leads show a few isolated
large peaks, separated by $15-25$ levels with much smaller peak
height (not visible on the scale of Fig.~1(b)). All large peaks are
associated with states quantized on stable bouncing--ball orbits.
This is illustrated in the inset for the state $p=337$.  The height
of the small peaks not resolved in Fig.~1(b) is typically two or
more orders of magnitude smaller than the maximum peak height. Such
tiny peaks are difficult to resolve in standard Coulomb blockade
experiments. The interference experiments \cite{Yac95,Schu97} face
additional noise from the transmission through the free arm of the
ring. For wide leads, interference experiments are therefore only
sensitive to the strongly coupled bouncing--ball modes.

The crucial role of the transverse barrier width $a_{\em eff}$ can
be understood from the following argument: Confinement in a lead of
width $a_{\rm eff}$ yields the transverse momentum spread $\hbar /
a_{\em eff}$ for electrons injected in the quantum dot.  Wide leads
therefore result in near normal injection and provide exceptionally
strong coupling to the bouncing--ball states. To quantify this
argument we express the partial width $\Gamma_p^l =|V_p^l|^2$ in
terms of the Husimi function $H_p$.  Recalling the Gaussian form of
$\psi_l$ and using the arc length $s$ as integration variable in
Eq.~(\ref{eq_over}), we find
\begin{eqnarray}
\label{eq_Husimi}
\Gamma_p^l =  \left( { |c_l| \hbar^2 \over m^*} \right)^2 H_p
(s^l, 0) ,
\end{eqnarray}
where $H_p$ is calculated for the normal derivative of the wave
function $\psi_p$. The Husimi function is evaluated at the phase
space point ($s = s^l,p_s = 0)$ reflecting the position $s_l$ of the
lead and zero transverse momentum. We note that the Husimi function
can be written as an integral over the Wigner function, smoothed
with Gaussians both in arc length and transverse momentum. Their
width is given by $a_{\rm eff} / \sqrt{2}$ and $\hbar / \sqrt{2}
a_{\rm eff}$, respectively. We express these widths in terms of the
uncertainties $\Delta \phi$, $\Delta \sin \chi$ in polar angle $\phi$
and injection angle $\chi$, respectively ($\chi$ is the angle between
the momentum of the injected electrons and the normal to the
boundary). Using the relation $p_s = p \sin \chi$, we find $\Delta
\phi \approx 0.7 (k a_{\rm eff}) / (k R_0)$ and $\Delta \sin \chi
\approx 0.7 /(k a_{\rm eff})$, where $R_0$ is the average radius of
the dot. 

The phase space portrait of the dot with deformation $\epsilon = 0.2$
is shown in Fig.~2 using $(\phi, \sin \chi)$ coordinates. 
Superimposed is the Husimi function of the bouncing--ball mode $p=337$
calculated for $k a_{\em eff}=5.0$ and $k R_0 \approx 40$,
corresponding to approximately $400$ electrons on the dot. The Husimi
function attains maximum value at the phase space points $(\phi,\sin
\chi) = \pm (\pi/2,0)$ representing normal injection from the leads.

In order to obtain the Husimi function and hence $\Gamma_p$
analytically, we must solve for the wave function $\psi_p$. For
states localized on a chain of stable islands the calculation can be
carried out using the semiclassical theory developed in
Refs.~\cite{NarSto99,NarHac99}. The result can be expressed in terms
of the stable periodic orbit at the center of the island,
\begin{eqnarray}
\label{eq_calc}
H_p(s,p_s)= & & \sum_\mu { A_{\mu} \over \Delta s_\mu \Delta p_\mu}
\exp\left[-\left(s-s_\mu \over \Delta s_\mu \right)^2 \right]  \nonumber \\
& & \hspace*{1cm} \times \exp\left[ - \left(p_s-p_\mu \over \Delta p_\mu 
\right)^2 \right] , 
\end{eqnarray}
where $s_\mu$ is the arc length and $p_{\mu}$ the transverse
momentum at the bounce points $\mu$ of the periodic orbit. Moreover,
$\Delta s_\mu = [a_{\rm eff}^2 + l_\mu^2]^{1/2}$ and $\Delta p_\mu=
[(\hbar /a_{\rm eff})^2 + (\hbar /l_\mu)^2]^{1/2}$ where $l_\mu =
\sqrt{2 \hbar |m_{12}^\mu|} |4-{\rm Tr}^2 [M_\mu] |^{-1/4}$ is
determined by the monodromy matrix $M_\mu \equiv (m_{ij}^\mu)$. The
amplitude $A_\mu$ depends only weakly on $a_{\rm eff}$. The result
(\ref{eq_calc}) allows us to express the width of island states in
terms of the phase space distance of the underlying periodic orbit
from the lead injection points. No comparable analytical result is
known for states quantized in the chaotic sea. However, we find
numerically that for wide leads the Husimi function of such states
has exponentially small support near the injection points (see the
inset of Fig.~2).

\begin{figure}[t] 
\begin{center} 
\leavevmode 
\vspace*{-1.1cm} 
\epsfxsize = 7.2cm 
\epsfbox{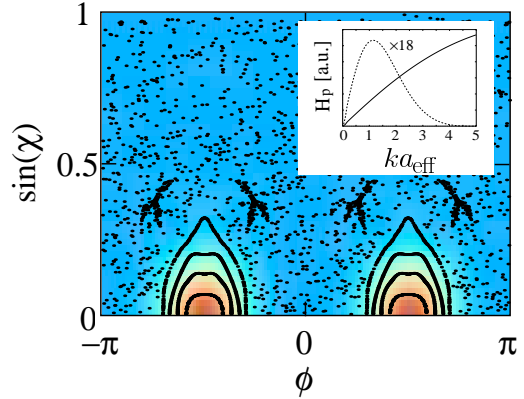} 
\end{center}
\vspace*{1.0cm} 
\caption{Husimi projection (color scale: red indicates high
  probability) of the bouncing--ball mode $p=337$ calculated for $k
  a_{\rm eff}=5.0$ and $\epsilon=0.2$. Superimposed is the Poincare
  surface of section of the classical phase space in $(\phi, \sin
  \chi)$ coordinates. Inset: $H_p (\phi=\pi/2,0)$ vs.\ $k a_{\rm
    eff}$ for the states $p=337$ (solid line) and $p=347$ (dashed
  line, enhanced by a factor $18$), the latter state is localized in
  the chaotic region of phase space.}
\end{figure} 

We now turn to the calculation of the transmission coefficient
$T_{\rm QD}$ and the phase $\phi_{\rm QD}$. We assume wide leads and
temperature $kT < \Delta$. The calculation for the weak coupling
limit $\Gamma_p \ll kT, \Delta$ is straightforward and can be done
using the Green function derived in Ref.~\cite{Bal99}; the results
will be presented elsewhere \cite{Hac00}. Here, we address the
regime of a more open dot as realized in the experiment
\cite{Schu97}. This regime is characterized by $\Gamma_{\bar{p}}
\sim \Delta$ where we identify $\bar{p}$ with the bouncing--ball
state closest to the Fermi energy.  All other states $p \neq
\bar{p}$ in the vicinity of $E_F$ have a width $\Gamma_p \ll
\Delta$. The Green function is found using the equations--of--motion
method \cite{Mei91}. It is diagonal up to small off--diagonal
corrections ${\cal O} (\sqrt{\Gamma_p \Gamma_{\bar{p}}} /\Delta)$,
and given by
\begin{eqnarray}
\label{Green}
G_{\bar{p} \bar{p}}(E) & = & \sum_{N=0}^\infty { P_{N}
\over  E - [E_{\bar{p}} -eV_g + U N] + i 
\Gamma_{\bar{p}}/2 }.
\end{eqnarray}
Here $N$ counts the total number of electrons in all dot levels {\em
  except} for the level $\bar{p}$, $P_{N}$ is the respective
occupation probability, $E_{\bar{p}}$ is the single--particle energy
of state $\bar{p}$, $V_g$ is the gate voltage, and $U=e^2/C$ is the
charging energy. The transmission through the states $p \neq \bar{p}$
is negligible. Equation (\ref{Green}) is the generalization of the
weak--coupling result \cite{Bal99} to the case of a single strongly
conducting quantum state.  The probability $P_{N} = Z^{-1}
\exp[-\Omega (N)/kT]$ with $Z=\sum_{N} \exp[-\Omega (N)/kT]$ is
related to the thermodynamic potential $\Omega(N)$ of the dot.  To
evaluate $P_{N}$ we replace $\Omega_{N}$ by $\Omega_{N}^0 +
[E_{\bar{p}} -eV_g +U N -E_F] \langle n_{\bar{p}} \rangle_{N}$, where
$\Omega_ {N}^0$ is calculated for the dot with level $\bar{p}$
excluded from the spectrum, and 
\begin{eqnarray} \langle
n_{\bar{p}} \rangle_{N} = -{1 \over \pi} \int \! dE  \, {\rm Im} {f(E)
\over E-[E_{\bar{p}} -eV_g + U N] +i \Gamma_{\bar{p}} /2}
\end{eqnarray} 
is the canonical occupation probability of level $\bar{p}$.

In Fig.~3 we show the transmission coefficient $T_{\rm QD}$ and the
phase $\phi_{\rm QD}$ vs.\ gate voltage $V_g$ calculated for $kT
=0.2 \Delta$, $\Gamma_{\bar{p}}=1.5 \Delta$ and $U=12 \Delta$. All
peaks shown result from transmission through the level $p=337$. The
peaks have comparable height and similar phase in qualitative
agreement with the experiments \cite{Yac95,Schu97}. The central peak
has a Lorentzian shape of width $\Gamma_{\bar{p}}$. Note that the
transmission peaks develop a peculiar asymmetry as the conducting
level moves away from the Fermi energy: Each peak to the left and to
the right of the central peak falls off more rapidly on the side
facing the central peak. This pattern extends over the whole
sequence and becomes more pronounced for the peaks in the tails.
{\em This asymmetry results from the breaking of the particle--hole
  symmetry in the transmission through the bouncing--ball state and
  is a unique fingerprint of BBT}. Inspection of the data of the
experiment \cite{Schu97} reveals the same asymmetry \cite{Goe99},
providing strong evidence that the peak correlations in this
experiment are due to BBT. We note that the number of correlated
peaks in Fig.~3 is less than observed in experiment. The origin for
this is addressed below. A detailed discussion of this issue and of
the phase lapse between the peaks will be given in a future
publication \cite{Hac00}.

\begin{figure}[t] 
\begin{center} 
\leavevmode 
\vspace*{-1.0cm}
\hspace*{-0.5cm} 
\epsfxsize = 7.4cm 
\epsfbox{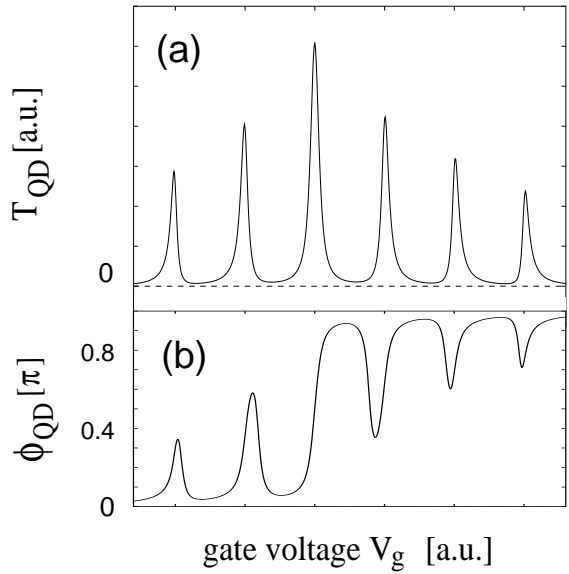} 
\end{center}
\vspace*{0.9cm} 
\caption{(a) Transmission coefficient $T_{\rm QD}$ and (b) phase 
  $\phi_{\rm QD}$ as a function of gate voltage $V_g$ evaluated for
  $kT = 0.2 \Delta$, $\Gamma_{\bar{p}}=1.5 \Delta$, and $U=12
  \Delta$. The peak asymmetry in (a) is a unique fingerprint of
  BBT.}
\end{figure} 

We finally address the universality of the results presented above.
The stability of bouncing--ball motion does not rely on the
effective dot potential considered here. Similar stability if found
for dots of other shapes \cite{Noe97,NarHac99} as well as for smooth
confining potentials \cite{Hac97}. We assumed that the dot shape
does not change upon the addition of electrons.  Some robustness of
the shape has indeed been demonstrated in a recent experiment
\cite{Ste97}. In general, however, variation of the gate voltage is
expected \cite{Hac97,Lew98} to change the electrostatic potential
and modify the shape of the dot. It was shown \cite{Hac97,Lew98}
that shape deformations can enhance peak correlations by ``pinning"
conducting levels close to the Fermi energy. To study the effect of
deformations, we varied the shape by a parameter linear in gate
voltage $V_g$. Over a range of $V_g$ we observed the predicted
enhancement of correlations and sequences of more than 10 correlated
peaks. At the same time, the bouncing--ball modes changed very
little with $V_g$ as substantial modification of the shortest stable
modes typically requires a large change in potential.

In conclusion, we have demonstrated a new mechanism for transport
through quantum dots in the Coulomb blockade regime. Current flows
by tunneling through bouncing--ball modes. The fingerprints of
bouncing--ball tunneling are sequences of correlated conductance peaks
with asymmetric peak line shape. We find the peak asymmetries in
recent Coulomb blockade interference experiments, providing strong
evidence that the peak correlations in these experiments are caused
by bouncing--ball tunneling.

We acknowledge helpful conversations with F.\ Haake and H.\ A.\
Weidenm\"uller. The work was supported by the Deutsche
Forschungsgemeinschaft, DGAPA--UNAM, and CONACyT--Mexico.


\begin{references}
\bibitem{shape} These dots were defined by surface gates of
  rectangular shape. The gates create a smooth dot shape in the
  electron gas below the surface.
\end{references}
\end{document}